\documentclass[journal]{IEEEtran}

\usepackage{cite}
\usepackage{amsmath,amssymb,amsfonts}
\usepackage{algorithmic}
\usepackage{graphicx}
\usepackage{balance}
\usepackage{textcomp}
\usepackage{wrapfig}
\usepackage{url}
\usepackage[detect-none]{siunitx}
\usepackage{tabularx}
\makeatletter
\g@addto@macro{\UrlBreaks}{\UrlOrds}
\makeatother
\def\BibTeX{{\rm B\kern-.05em{\sc i\kern-.025em b}\kern-.08em
    T\kern-.1667em\lower.7ex\hbox{E}\kern-.125emX}}
\begin{document}
\bstctlcite{IEEEexample:BSTcontrol}
\title{Structural Health Monitoring system with Narrowband IoT and MEMS sensors}
\author{Flavio Di Nuzzo, Davide Brunelli, \IEEEmembership{Senior Member, IEEE}, Tommaso Polonelli, \IEEEmembership{Student Member, IEEE},\\ and Luca Benini, \IEEEmembership{Fellow, IEEE}
\thanks{This work was supported by Sacertis S.r.l. and by the Italian Ministry for Education, University and Research (MIUR) under the program “Dipartimenti di Eccellenza (2018-2022)”. }
\thanks{Moreover this research is funded by ECSEL, the Electronic Components and Systems for European Leadership Joint Undertaking under grant agreement No 826452 (Arrowhead Tools), supported by the European Union Horizon 2020 research and innovation programme and by the member states. }
\thanks{F. Di Nuzzo, T. Polonelli and L. Benini are with the Department of Electrical, Electronic and Information Engineering, University of Bologna, 40136 Bologna, Italy.are with University of Bologna, Bologna, Italy (e-mail: \{name.surname\}@unibo.it). }
\thanks{D. Brunelli is with the Department of Industrial engineering, University of Trento, 38123 Trento, Italy (e-mail: davide.brunelli@unitn.it).}
\thanks{L. Benini is also with the Department of Information Technology and Electrical Engineering at the ETH Zurich, 8092 Zurich, Switzerland. (e-mail: lbenini@iis.ee.ethz.ch).}
\thanks{
This article has been accepted for publication in a future issue of this journal, but has not been fully edited. Content may change prior to final publication. Citation information: DOI 10.1109/JSEN.2021.3075093, IEEESensors Journal}
\thanks{Personal use is permitted, but republication/redistribution requires IEEE permission. See \url{http://www.ieee.org/publications_standards/publications/rights/index.html} for more information.}
}

\maketitle
\begin{abstract}
Monitoring of civil infrastructures is critically needed to track aging, damages and ultimately to prevent severe failures which can endanger many lives.  The ability to monitor in a continuous and fine-grained fashion the integrity of a wide variety of buildings, referred to as structural health monitoring, with low-cost, long-term and continuous measurements is essential from both an economic and a life-safety standpoint. 
To address these needs, we propose a low-cost wireless sensor node specifically designed to support modal analysis over extended periods of time with long-range connectivity at low power consumption. Our design uses very cost-effective MEMS accelerometers and exploits the Narrowband IoT protocol (NB-IoT) to establish long-distance connection with 4G infrastructure networks. Long-range wireless connectivity, cabling-free installation and multi-year lifetime are a unique combination of features, not available, to the best of our knowledge, in any commercial or research device. We discuss in detail the hardware architecture and power management of the node. Experimental tests demonstrate a lifetime of more than ten years with a 17000~mAh battery or completely energy-neutral operation with a small solar panel (60~mm x 120~mm). Further, we validate measurement accuracy and confirm the feasibility of modal analysis with the MEMS sensors: compared with a high-precision instrument based on a piezoelectric transducer, our sensor node achieves a maximum difference of 0.08\% at a small fraction of the cost and power consumption.
\end{abstract}



\section{Introduction}
\label{sec:introduction}
In our cities, buildings and civil infrastructures are increasing every year~\cite{romano2017land}. New materials and techniques applied to complex structures create the need for continuous and autonomous monitoring of the structural integrity of the buildings over many years~\cite{romano2017land, calvi2019once}. Furthermore, old structures and facilities such as bridges, viaducts, roads, and apartments have to be continuously checked to guarantee the safety of thousands of people: in fact, old civil and mechanical engineering structures continue to be used daily despite aging, deterioration, and exceedance of their operational lifespan. For instance, between 2013 and 2018,  one out of nine of the United States bridges were structurally deficient, while in Italy, 13 bridges collapsed, in addition to a 210 meters highway bridge breakdown, which killed 43 people~\cite{calvi2019once}. 
According to these facts, the ability to monitor a wide variety of buildings with low-cost and real-time measurements is essential from both an economic and a life-safety standpoint. 
The process of continuously monitoring the integrity and the response of a structure is referred to as Structural Health Monitoring (SHM)~\cite{tokognon2017structural}. 
The primary function of an SHM system is to provide a continuous assessment of structural integrity and promptly identify, or even predict, damages and abnormal structure behaviors while generating periodic reports about the structure state~\cite{tokognon2017structural}.
In particular, civil engineers and inspectors need periodic data to assess the mechanical performance over long time intervals, typically many years, and to analyze the structure’s long-term response to environmental and daily stresses (i.e., weather, wind, weight load, corrosion, or act of vandalism). Both static and dynamic measurements are required to achieve this goal: corrosion and crack monitoring are typical static measurements, while modal analysis~\cite{shm_zonzini} 
needs dynamic (velocity, acceleration) measurements.
A SHM system comprises four blocks: i) sensors and transducers, ii) remote communication, iii) data storage, and iv) feature extraction and data processing. Depending upon the SHM installation complexity and operation context, the system could work in small scale, for example, on a single structure or exploiting many thousands of sensors with national coverage. In this paper, we focus our effort on the second case, providing support for sparse installation of heterogeneous IoT devices for national or internationals SHM networks. 
Historically, SHM devices were produced using wired networks; however, the low installation and maintenance costs and high reliability of state-of-the-art LPWAN (Low-Power Wireless Area Network) and Internet of Things (IoT) protocols have made them a compelling alternative solution for industrial deployments~\cite{dhage2017structural,valenti2018low}.  
Indeed, due to their high installation costs, wired systems are generally available only for long-term monitoring of important key structures. SHM wireless sensors enable significant cost reductions~\cite{catenazzo2018use,Burrello_iot}, allowing a massive utilization in public and private infrastructures.  Moreover, such sensors are also suitable for short-term structural monitoring applications, exploiting single clusters of devices for multiple buildings. 
Previous works have pointed out that, concerning power consumption, coverage, and security, LoRaWAN~\cite{haxhibeqiri2018survey} and NB-IoT~\cite{ballerini2019experimental,ballerini2020nb} are the most suited LPWAN protocols in distributed IoT applications.
In addition to the installation, the second limiting factor comes from transducers~\cite{sabato2016wireless}. Indeed, it is not obvious which kind of behavior better describes the structure health, and the sensor cost is itself a major limiting factor. Ordinarily, to read the small vibrations and oscillation frequency components of a building with sufficient precision ($10^{-2}m\cdot s^{-2}$)~\cite{sabato2016wireless}, piezoelectric monoaxial sensors are used. 
Furthermore, to characterize a wall crack through time and space, LVDT (Linear Variable Differential Transformers), optical-fiber based sensors~\cite{wang2019strain}, and laser sensors are often exploited, which a sensitivity below one micrometer.
They are usually noted for their high cost, characterized by the transducer and the complex analog front-end. Moreover, the harness's deployment to connect several devices in a large and external environment complicates the installation and raises the costs. Due to these facts, SHM systems are not commonplace and widespread as they should~\cite{popkova2019preconditions}.
To overcome these limitations, previous works proposed the integration of Micro-Electro-Mechanical Systems (MEMS) in SHM devices for making low-cost monitoring systems~\cite{sabato2016wireless,shm_brunelli}. Currently, structural instrumentation using MEMS provides low-cost installation, moderate invasive effects, and equivalent performance compared to their macro-scale counterparts~\cite{sabato2016wireless}.  Off-the-shelf MEMS integrated System on Chips (SoCs) are appealing for SHM application as they measure vibrations from $10^{1}m\cdot s^{-2}$ (severe shaking) to $10^{-2}m\cdot s^{-2}$ (micro-vibration) on large-scale structures having natural oscillation frequencies in the range of $10^{-1}$ to $10^{1}$Hz~\cite{naeim2007dynamics}.  
Moreover, it has been demonstrated that data from MEMS inclinometers can also provide additional
information about the structural health of the system with respect to an increasing traffic load~\cite{burrello2020enhancing}.
This work presents a low-cost wireless SHM monitoring sensor board that uses low-cost MEMS sensors and data communication using the NB-IoT protocol.
It is specifically designed to support building modal analysis through low-cost (price between 1.83€ and 2.17€ at 1kunit) commercial MEMS sensors to facilitate wide-scale SHM systems' deployments. Using NB-IoT communication, the sensor board guarantees extremely low power consumption ($125.4~mW$ run, $145.2~\mu W$ sleep) and capable of global (nation-scale) connectivity through the NB-IoT infrastructure network. 
A combination of onboard processing and data aggregation decreases the communication energy, $50\mu J$ per bit on average, by exploiting the floating-point unit (FPU) performance of the STML4 MCU series
Indeed, applying the edge-computing methodology in our sensor board enables a data compression rate of 256. For high-accuracy and short term deployment using a non-rechargeable thionyl chloride 17000~mAh battery, our device reaches 214 days of operativity. Moreover, for long-term monitoring using the same battery, the 10 years threshold is achieved, considering 6 samples per day, in which 3 MB of vibration values compose each sample. The node has also been tested with onboard solar energy harvester support self-sustainability using a $60~mm\times 120~mm$ solar panel and a Li-On battery, used as an energy buffer, of $5400~mAh$. 
The measurement precision of our device is comparable with piezoelectric sensors. To achieve this result, we performed a comprehensive MEMS assessment to select the best commercial product in comparison with high precision piezoelectric sensor. Our device measurements perfectly match results from expensive and power-hungry piezoelectric sensors, providing an error lower than 0.08\% in frequency peak detection of modal vibration tones.
The rest of the paper is organized as follows. 
Related work is reviewed in Section~\ref{sec:related}. Section~\ref{sec:background} presents a design exploration of NB-IoT, providing consumption analysis and discussion of the limits. Performance comparison of three commercial modules has been discussed to justify the more suitable communication module for the designed SHM node. We present the hardware and software design of our SHM board in Section~\ref{sec:sensor}. An energy consumption model is elaborated in Section~\ref{sec:e_model}, and it is also validated with laboratory measurements. Section~\ref{sec:in_field_valid} describes all laboratory simulations and results to test and evaluate the board's performance compared to state-of-art sensors. Section~\ref{sec:concl} concludes the paper.

\section{Related Work}
\label{sec:related}
Structural Health Monitoring (SHM) has been studied for some decades already; as we can see in \cite{shm_1999} studying the modal vibration of a structure, it is possible to detect damages in advance. Furthermore, more recent studies developed approaches of damage detection based on autoregressive (AR) time series and damage sensitive features (DSFs) defined as a function of the AR coefficients \cite{shm_damage_detec}. Another example is given by \cite{SHM_algo} where the AR coefficients are calculated with autoregressive moving average (ARMA) processes and then given to a Gaussian Mixture Model (GMM) to model the feature vector. These two algorithms permit to detect damage, but also to locate it in the structure and, most importantly, they use vibration data coming from inertial sensors. Considering the improvements in microcontrollers' computational performance and considering the last generation of MEMS with higher precision and newer features of machine learning, it's possible to perform SHM in an embedded system. 
A major trend in SHM is to replace expensive piezoelectric sensors with more affordable MEMS sensors. In~\cite{shm_demarchi}, a comparison between a piezoelectric sensor and a MEMS sensor for vibration-based SHM is presented, demonstrating that MEMS is a viable technology for SHM with a remarkable cost reduction. Another example is given by~\cite{shm_brunelli},which has been validated in production and serves as the initial reference for this paper.  The main limitations of this design are: (\textit{i}) it uses short-range wireless communication for data (Wi-Fi) that is also power-hungry (0.5~W); (\textit{ii})it is not designed for battery operation, and requires power cabling.

In~\cite{polonelli2018crackmeter}, a sensor node that measures and monitors cracks in concrete for SHM purposes is proposed.
The sensor node uses the LoRaWAN protocol to communicate data, and it guarantees a battery lifetime of more than 10 years.
Highly accurate static measurements of crack's amplitude are demonstrated along with Long-range connectivity.
However, the bandwidth and sample rate of crack metering is orders of magnitude lower than what is needed for vibration-based SHM.
Concerning static measurements~\cite{wang2019strain,wang2018strain2,polonelli2018crackmeter}, for example, cracks width or corrosion, it is possible to detect damages in structures using strain gauges. These measurements have low bandwidth and can be performed with very low power sensor and electronics, which can survive in the building for all its lifetime. Our work focuses on much higher bandwidth vibration-based analysis, which is as essential for SHM as static analysis since it provides complementary information~\cite{shm_brunelli,shm_demarchi,ragam2019shm}. Ideally, dynamic analysis should be performed in a continuous, long term fashion, in parallel with static analysis, but it generally is not, because of power and cost concerns.
The designs proposed so far in the literature and in the industrial practice are not suitable for vibration-based (dynamic), battery-powered SHM with long lifetime. 
In this work, we improve over previous designs in several directions. First, we use the NB-IoT protocol for our communication link and leverage its advantages over LoRaWAN ~\cite{polonelli2018crackmeter}: NB-IoT uses a licensed band, so there is no limitation on band occupation; moreover, there is no need of gateways because the node connects directly with mobile operator's base stations. Note that, LoRaWAN protocol is generally less energy consuming than NB-IoT~\cite{ballerini2019experimental} at low datarates, but it is not adequate for high datarate SHM applications that need thousand of KBytes of data transferred. The NB-IoT payload is 3 times bigger than LoRaWAN (1500 Bytes against 300 Bytes in specific condition); furthermore, NB-IoT datarate is $200~kbps$ against $50~kbps$ of LoRaWAN \cite{mekki2019lpwan}. Considering a long term application as in this work, 12kB per inertial measurement session must be sent, for a total of 6 sessions. 72kB per day is unfeasible using LoRaWAN protocol because of the low data rate and, more importantly, for the regulated band occupation that only permits very low duty-cycle. NB-IoT has no limitation on data transmission, and the energy per bit (EPB) with big payload ($>$5kB) is affordable, as it is lower than $50\mu J$ (as shown in Table~\ref{epb_tab} in Section~\ref{sec:sensor}).

After evaluating the performance of MEMS analog sensor presented in~\cite{sabato2016wireless}, we decided to use in our work the LIS344ALH
by ST Microelectronics because of its leading edge characteristics among MEMS sensors. 
In fact, LIS344ALH has a noise density of $50~\mu g/\sqrt{Hz}$ (the lowest value in this class of devices), and its performance is confirmed in \cite{shm_brunelli}. Moreover, thanks to the efficiency of the microcontroller STM32 series L4 from ST Microelectronics, we implemented an acquisition chain designed to oversample and filter the data coming from the sensor, achieving 16~bit ADC resolution.
To the best of our knowledge, there are no other devices that exploit the NB-IoT protocol to enable vibration-based SHM. In literature,\cite{alonso2018shmwsnsurvey,arcadius2017shmwsnreview} provides several examples and evaluations on wireless communication applied to SHM, but none of them use NB-IoT. Also, in~\cite{ragam2019shm} are presented some examples of SHM systems based on MEMS sensors, but in all proposed solutions, the usage of NB-IoT is not considered. 
\section{Background}
\label{sec:background}
\subsection{Narrowband IoT}
\begin{figure}[!t]
\centering
\includegraphics[width=\columnwidth]{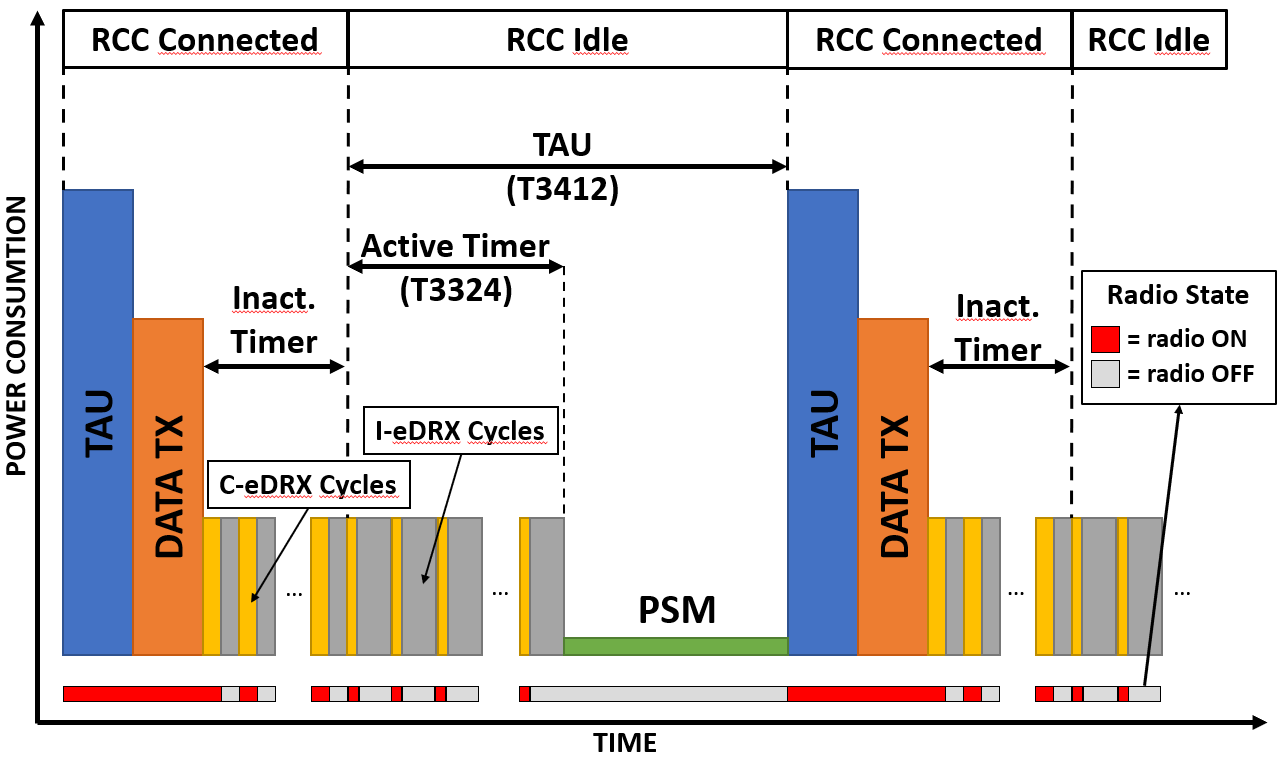}
\caption{Behaviour of a NB-IoT module with description}
\label{nb_datagram}
\end{figure}

The NB-IoT is a Low Power extension of the LTE (4G Long Term Evolution) developed for long battery lifetime and low cost application. Its main features are reduced power consumption, extended coverage extension, user equipment cost reduction, and backward compatibility~\cite{wanglin}. The NB-IoT's power consumption depends on multiple environment-related factors, such as the country and network operator settings, which can drastically change the end-device performance. NB-IoT is standardized by 3GPP for LPWANs thanks to its capability to work virtually everywhere (e.g., TIM guarantees a country level coverage in Italy for NB-IoT)~\cite{ballerini2020nb}. 
The maximum payload for each message is 1500~B, and the full data transmission rate is limited to $20~kbps$ uplink and $200~kbps$ downlink. The minimum bandwidth is 180~kHz, corresponding to the size of the smallest LTE Physical Resource Block (PRB). As presented in~\cite{perfNBiot}, this protocol is meant for extended battery life applications of more than ten years when transmitting 200 bytes per day, features that fit into classic SHM requirements. Low power consumption is achieved by NB-IoT using the LTE energy-saving mechanisms, extending the inactivity periods to minimize energy consumption. 
As reported in~\cite{nb-boundaries}, the RCC state model (LTE Radio Resource Control) protocol has only two states: RCC Connected and RCC Idle (Fig.~\ref{nb_datagram} - RCC), the cell handover and redirection is not supported in NB-IoT release 13. 
NB-IoT features a Discontinuous Reception (DRX) mechanism, where the module alternates active listening for Paging Occasions and sleep periods. DRX settings are defined in multiple subframes of 1~ms, ranging from 256~ms to 9216~ms against 10~ms to 2560~ms of LTE to reduce power consumption. NB-IoT defines it enhanced DRX in RRC Connected mode (C-eDRX) (Fig.~\ref{nb_datagram} - C-eDRX Cycles). In RCC Idle, the eDRX in Idle state (I-eDRX) (Fig.~\ref{nb_datagram} - I-eDRX Cycles) is equal to the one in LTE, but also, in this case, the timings can be larger 
It can return in RCC Connected state at any time, using the same context saved during the initialization (first connection to the mobile operator's cell). The reconnection process from PSM state to RCC connected state decreases latency and energy consumption compared to establishing a new connection in the initialization phase. 
T3412 and T3324 set the duration of PSM and I-eDRX, respectively (Extended Timer and the Active Timer (Fig.~\ref{nb_datagram} - T3412 and T3324).  The Active Timer (T3324) determines the duration of the I-eDRX after entering RCC Idle state, and the Extended Timer (T3412) sets the period of the Tracking Area Update (TAU). The TAU is the same as that of LTE, but NB-IoT can also configure a longer period of up to 413 days (spent in PSM).

\subsection{NB-IoT module: a comparative study}
\begin{figure}[!t]
\centering
\includegraphics[width=\columnwidth]{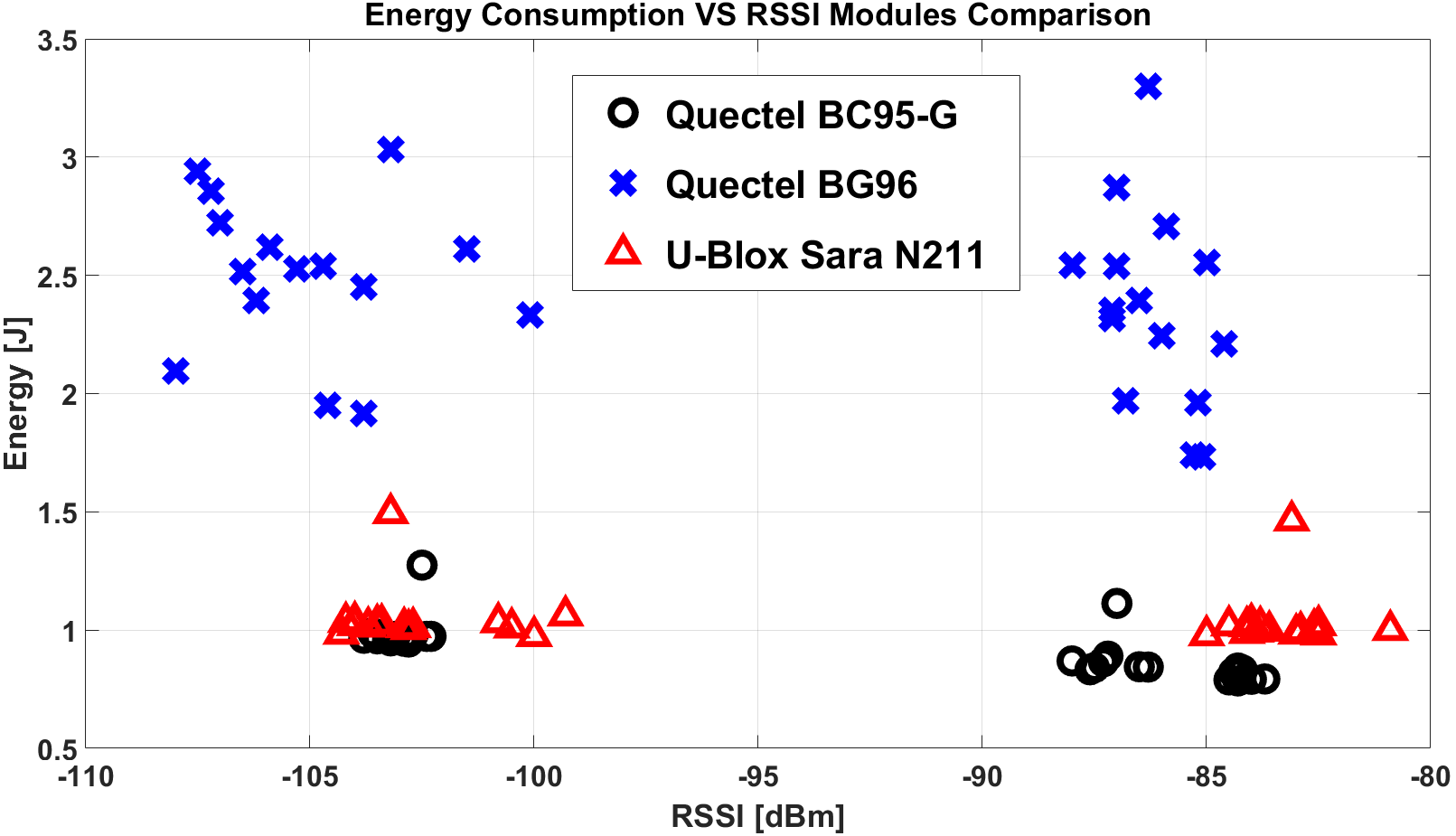}
\caption{Power Consumption in transmission: Quectel BC95, Quectel BG96 and U-Blox Sara N211}
\label{module_comp}
\end{figure}

We compared three different modules available on the market: the Quectel BG96, the Quectel BC95-G and, the U-Blox Sara N211. The Sara N211 module is mounted on the development board C030-N211 produced by U-Blox; the modules from Quectel are mounted on development boards build by MIKROE, called  MIKROE-3144 and MIKROE-3294, respectively for the BG96 and the BC95-G. 
We mainly focused on each module's energy consumption in PSM and the energy per packet required for each uplink in different coverage conditions.
The U-Blox dev-board also mounts a current measuring circuit, consisting of a shunt resistor placed on the supply voltage and a differential amplifier based on the TSZ121
operational amplifier. For a fair comparison, we built the same current measurement setup on the MIKROE dev-boards. 
We sent the same packet using UDP protocol and measured the energy, latency, and reliability. 
Moreover, for each transmission, we assessed the RSSI (Received Signal Strength Indicator) to classify the level of coverage. Using a -20~dB coaxial passive attenuator, we simulate different connection conditions. 
Results in Fig.~\ref{module_comp} gather all the acquired data. The vertical axis represents the power consumption expressed in Joule, and the horizontal one shows the RSSI in dBm.
Taking as a reference Fig.~\ref{module_comp}, the following considerations are validated. The Quectel BG96 has higher energy consumption; indeed, it consumes 71\% more on average compared to the Quectel BC95-G and the U-Blox Sara N211; this fact is given by the extra functionalities of this module. 
The Quectel BG96 can also operate as an LTE-M module or GPRS (General Packet Radio Service) module; furthermore, it has additional interfaces and even a GNSS (Global Navigation Satellite System) module.
The U-Blox Sara N211 and the Quectel BC95-G are comparable; in good coverage, the Quectel BC95-G consumes 15\% less energy on average than the U-Blox Sara N211; in medium coverage, the energy consumption is almost the same. 
We selected the Quectel BC95-G instead of the BG96 because of the lower average energy consumption and for the NB-IoT focused firmware. In fact,  the latter provides a set of proprietary AT commands with extended functionalities not needed in our application scenario, such as commands to configure and use the MQTT (Message Queue Telemetry Transport) protocol or 
to enable LTE and GNSS functionalities. 

\subsection{Quectel BC95-G characterization and energy considerations} 
\label{ssec:BG95}
We measured the energy consumption of the BC95-G in different conditions. Moreover, aiming to design a reliable and plug \& play device, we assessed NB-IoT technology boundaries in SHM like environments.
We perform more than 600 measurements using different passive and active RF  attenuator to simulate different coverage situations. Measurements are reported in Fig.~\ref{bc95vsrssi}. We grouped the measurements in 3 areas defined by the RSSI:

\begingroup
\setlength{\tabcolsep}{1pt}
\begin{tabular}{l r c l}
\\ 
  Good Coverage: &$-95~dBm<$&RSSI;& \\ 
  Medium Coverage: &$-110~dBm <$&RSSI&$< -95~dBm$;\\
  Bad Coverage: &&RSSI&$< -110~dBm$; \\
\end{tabular}
\\ 
In the following tests, the RSSI spans from a maximum of $-75~dBm$ to a minimum of $-120~dBm$, a typical received power in most real application scenarios.
In Fig.~\ref{bc95vsrssi}, the energy consumption is lower than $2~J$ (with one outlier) for transmission with RSSI greater than $-96~dBm$, the mean energy consumption is lower than $1.1~J$ with RSSI greater than $-96~dBm$.
In case of an RSSI lower than $-110~dBm$ the energy consumption is $4.071~J$ in average because of the bad coverage level, and ECL (the Enhanced Coverage Level) is equal to 2. Hence the module performs multiple ($2^i$ with $i=1...7$ based on ECL~\cite{nb-boundaries}) re-transmission to send the packet correctly. 
In Bad connectivity conditions, each packet requires $3.8\times$ more energy than in Good or $2.8\times$ more than in Medium conditions; this is a significant difference which must be kept into consideration while estimating the battery life.

\begin{figure}[!t]
\centering
\includegraphics[width=\columnwidth]{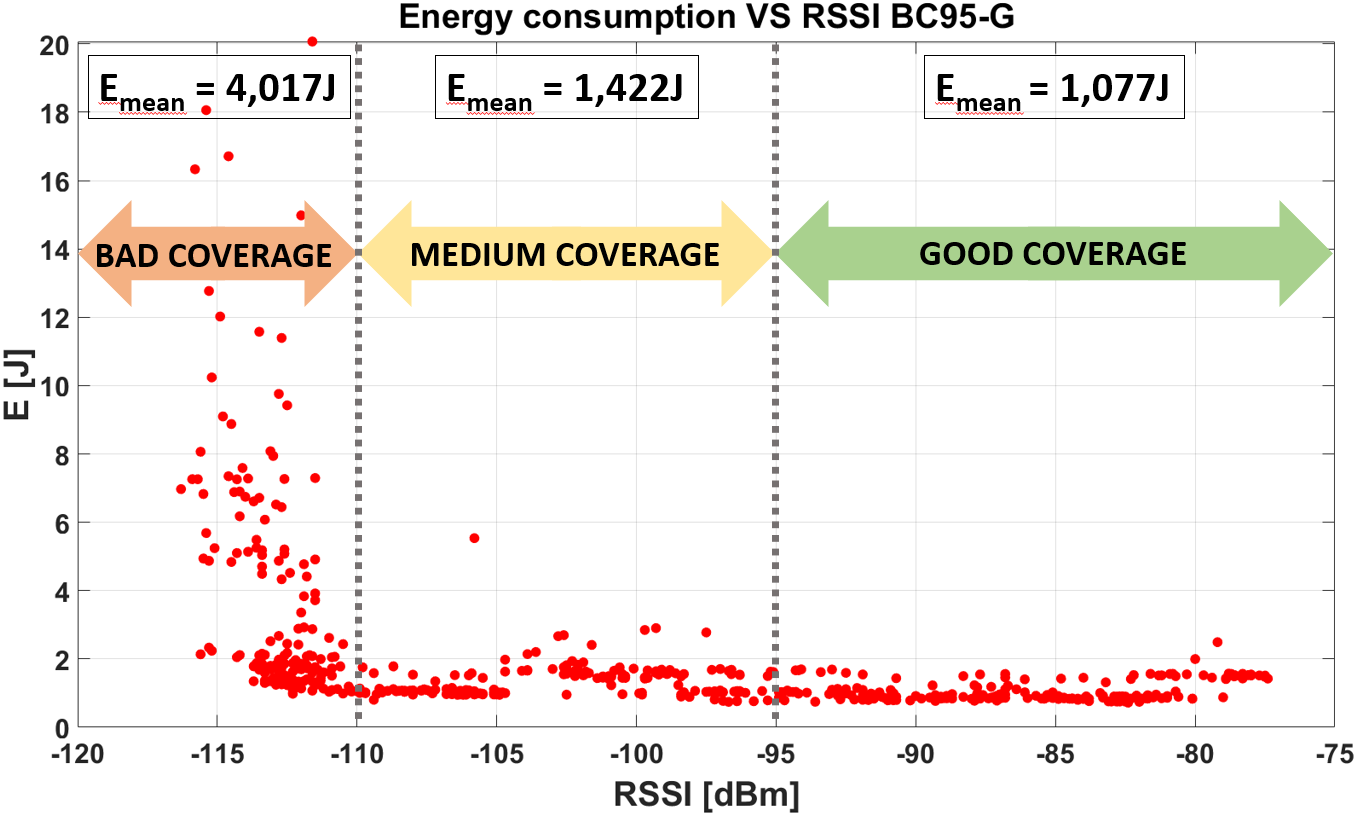}
\caption{Quectel BC95-G: average energy consumption in different coverage levels}
\label{bc95vsrssi}
\end{figure} 

In comparison with LoRaWAN, or other LPWAN protocols, the NB-IoT requires more energy per packet in general. However, due to its larger payload and its national licensed connectivity, the energy per bit from the sensor to the cloud is the best in class for the LPWAN category~\cite{ballerini2020nb}.

NB-IoT protocol has no restriction on band utilization because it uses licensed LTE bands, so there is no limitation on the number of bytes sent in a single connection to the cell. In  SHM applications, NB-IoT is ideal because gathered data can be buffered, and latency requirements are not so severe. Several energy consumption measurements with different payloads size have been done. Table~\ref{epb_tab} reports the mean energy consumed for each payload size and the energy per bit (EPB). Notice that the EPB with $500~Bytes$ of data transmitted is $235.1~\mu J$, thus 5.53 times more than the EPB with $10.8~KBytes$ of data. An SHM node can generate big chunks of data (hundreds of kBytes), that NB-IoT protocol can send with no restriction keeping a low power profile and with low EPB when data is buffered and sent in a single connection.

\begin{table}
\caption{Mean energy consumption and EPB with different payload sizes, the $\Delta$ difference is compared to 500~Bytes of Payload}
\centering
\footnotesize
\label{epb_tab}
\begin{tabular}{ccccc}
\hline\hline
\textbf{Payload}	& \textbf{Mean E.} & \textbf{$\Delta$}	& \textbf{EPB} & \textbf{\textbf{EPB}$\Delta$}\\
\textbf{[B]}	& \textbf{Consumption [J]} & 	& \textbf{[$\mu J$]} & \\
\hline
10		& 0.7130 & $\times1.32$			& 8912 & $\times0.02$\\
200		& 0.8123 & $\times1.15$			& 507.7 & $\times0.46$\\
500		& 0.9405 & ------			& 235.1 & ------\\
1300		& 1.0326 & $\times0.91$			& 99.29 & $\times2.36$\\
5400		& 2.1199 & $\times0.44$			& 49.07 & $\times4.79$\\
10800		& 3.6702 & $\times0.25$			& 42.48 & $\times5.53$\\
\hline\hline
\end{tabular}
\end{table}

\section{Sensor node design}
\label{sec:sensor}
\subsection{Hardware design}
After the exploration of the most suitable NB-IoT module and configuration, the hardware design of the complete SHM node is discussed in this section.
Recent works in the literature have already investigated the sensor front-end~\cite{shm_brunelli}, and the wireless connectivity~\cite{ballerini2020nb},  providing fundamental hardware specifications.
In fact, as the major problems of commercial SHM systems are the high sensor and installation cost, we combined MEMS sensor micro-vibration requirements~\cite{shm_brunelli} with an LPWAN transceiver~\cite{ballerini2020nb}. An SHM board with wireless connectivity and battery supply drastically reduces the installation costs. 
The main components of the designed prototype are:
\begin{itemize}
    \item \textbf{STM32L476} microcontroller by ST Microelectronics: L4 Series microcontroller with ARM Cortex-M4 and Floating Point Unit (FPU). It feature low power consumption (39~uA/MHz) and a direct SD card interface;
    \item \textbf{LIS344ALH} MEMS inertial sensor: high performance 3-axis $\pm$2~g/$\pm$6~g ultra-compact linear accelerometer; 
    \item \textbf{IIS2ICLX} High-accuracy, high-resolution, low-power, 2-axis digital inclinometer; it embeds a machine learning core for pattern recognition and automatic trigger generations. Moreover, it could provide inclinometric measures with an accuracy of 0.015~$mg/LSB$;
  \item \textbf{Quectel BC95-G} LTE-CatNB1 NB-IoT module, characterized in Section~\ref{ssec:BG95}
\end{itemize}

Furthermore, the board is equipped with two temperature sensors for onboard and external temperatures. A linear voltage regulator has been added to provide a stable supply voltage starting from the battery output. A USB compatible battery charger system with an integrated power switch for Li-Ion/Li-Polymer is mounted in case rechargeable batteries or an energy harvester are used. 
Two current measurement circuits are implemented in the SHM sensor node to sample the current used by the circuit in real-time. Measures are split into two power domains: the Quectel BC95-G and the rest of the components. Analyzing the entire board's power consumption permits the detection of abnormal current drains and, more importantly, the prediction of the residual battery energy. 
Indeed, as reported in the previous section, the physical positioning and the wireless connectivity could vary the power consumption of the communication link up to $3.8\times$.

The inclinometer IIS2CLX is used for two main tasks: gather clinometric data and provide an interrupt in case of a peak detection. To reduce power consumption
between two measurement sessions, the board is halted in a "sleep" condition, but in case of an extraordinary event (e.g., accelerometric peak caused by an earthquake), the IIS2CLX will wake-up to transmit emergency measurements. 
The final 4-layer Printed Circuit Board (PCB) is $120~mm\times 60~mm$ and is shown in Figure~\ref{set_up}.
\subsection{Firmware design and data acquisition}
The board Firmware is based on FreeRTOS to ensure code efficiency and scalability. Three main tasks are defined: 
\begin{itemize}
    \item \textbf{BC95-G managment Task}: it manages all the communications and network operations between the NB-IoT module and the microcontroller. The STM32 controls initial settings and the first connection procedure, than it enables both data transmission and reception;
    \item \textbf{SD card Task}: in this task, the MCU manages the data logging on the SD card. Measurements coming from different sensors and sessions are organized and saved in text files, providing a ready to use format for end-users;
    \item \textbf{Sensor acquisition Task}: in this task, all procedures that concern the sensors data acquisition are considered; in particular, the MCU handles acquisition and filtering of inertial sensor data.
\end{itemize}
In BC95-G management task, the microcontroller managed the NB-IoT communication providing suitable AT commands through a UART serial port. 
Furthermore, the state machine also includes a full set of recovery capabilities in case of errors or unexpected situations. For example, when the connection to the network operator stales or takes more than 1 minute, the software automatically resets the module and tries a new connection. 
During the transmission phase, the task manages the packed sending, supporting data fragmentation and replication for large volume uplinks. Afterward, it manages the PSM state and the RTC alarm for the next measuring sessions.
Concernign the sensor data acquisition, a filtering task is implemented. SHM typically measures  vibrations in the \numrange{0}{20}Hz bandwidth, while principal modes of the structure are in the range of \numrange{0}{10}Hz. On extraordinary stressful events for the structure (like an earthquake) the typical frequencies can reach a higher range, typically up to 45Hz~\cite{shm_brunelli}.
For this reason, the SHM node provides 16~bit acceleration values on the three axes at 100Hz data rate. 
We applied oversampling on the 12-bit ADC of the microcontroller to achieve 16-bit as ENOB (Effective Number Of Bits) necessary for achieving sensitivity and precision requirements. 
A further advantage associated with the oversampling technique concerns the post processing, which eases the implementation of digital filters. In fact, the digital FIR (Finite Impulse Response) filter is directly optimized for ARM Cortex-M4 to decimate data while applying a low pass filtration. The acquisition chain samples the data at $f_{OS}=25.6~kHz$ from the sensor, the gathered data is then processed by a FIR filter with a ratio decimation of 256, generating filtered data with 16 bit resolution at the final data rate of $f_{S}=100~Hz$.
The filter is a FIR filter composed by 6 different stages with a resulting decimation factor of 256 and a cutoff frequency of $50~Hz$. The minimum attenuation in stop band is $60~dB$ and the ripple in pass band is $0.1~dB$ peak-to-peak, maximum. The overall number of coefficients is 776, with a measured computation time of about $510~\mu s$, way lower than $10~ms$ ($f_{S}=100~Hz$).
The SD card task manages the session data logging. While the sensor acquisition task is running, the packets with filtered data are saved on an SD card in a text file format. 
A double buffering technique is used between this task and  the 
sensor task.  
\section{Energy Consumption model}
\label{sec:e_model}
\subsection{Battery Life Estimation}
Different current measurements in multiple operating conditions have been carried out to characterize the behavior of the SHM sensor. To estimate the battery lifetime, we have analyzed the power profile of a complete acquisition, composed by $5.2~kB$ of data. As shown in Fig.~\ref{sett_cons}, the energy consumption is partitioned into two major blocks: the energy used by data acquisition ($E_{acq}$) and the one consumed for data transmission ($E_{TX}$). These two blocks contribute as reported in Fig.~\ref{sett_cons}:

\begin{enumerate}
    \item $E_{acq1s}$: one second of acquisition at 100~Hz;
    \item $E_{SD-WR}$: write data on the SD Card;
    \item $E_{c-1TX}$: connect and transmit the first packet;
    \item $E_{pkt-TX}$: transmit one packet (general);
    \item $E_{cdrx-disc}$: energy consumed by Connected\_DRX and disconnect phases.
\end{enumerate}
\begin{figure}[!t]
\centering
\includegraphics[width=\columnwidth]{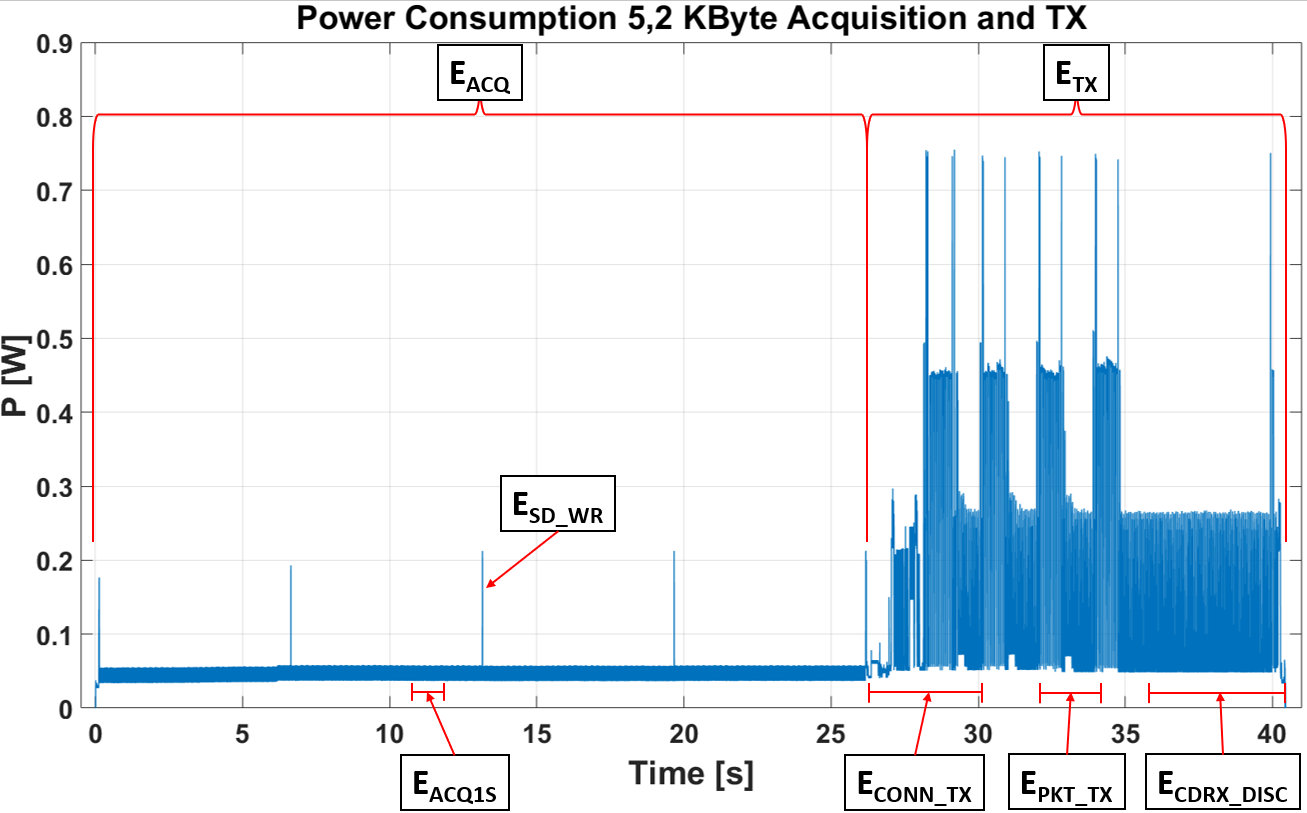}
\caption{Different energy contributions aligned on a measured transmission}
\label{sett_cons}
\end{figure} 
For each  acquisition and transmission phase, the NB-IoT module sends 1300~B, each packet consists of 650 filtered samples (16-bit depth per sample) saved on the SD card and thereafter transmitted. In Fig.~\ref{sett_cons}, the energy consumption varies with the number of packets ($N_{pkt-TX}$) given by Eq.~\ref{num_pkt}:
\begin{equation}
\footnotesize
    N_{pkt-TX}=\frac{T_{acq}*f_{s}}{Payload}.
\label{num_pkt}
\end{equation} 
where $T_{acq}$ is the time duration in seconds of the sampling time, and it's set by the user according to the application.
The energy consumption of the $E_{acq}$ is given by the Eq.~(\ref{acq_e}), the $E_{TX}$ is given by Eq.~(\ref{tx_e}) and the total energy consumption for a session is given by Eq.~(\ref{e_tot}): 
\begin{equation}
\footnotesize
    E_{acq}=E_{acq1s}(6.5*N_{pkt-TX})+N_{pkt-TX}*E_{SD-WR}
    \label{acq_e}
\end{equation}
\begin{equation}
\footnotesize
    E_{TX}=E_{c-1TX}+E_{cdrx-disc}+(N_{pkt-TX}-1)E_{pkt-TX}
    \label{tx_e}
\end{equation}
\begin{equation}
\footnotesize
    E_{tot}=E_{TX}+E_{acq}
    \label{e_tot}
\end{equation}
Table~\ref{e_cont} reports the energy consumption of every sub-block, 
highlighted in Fig.~\ref{sett_cons}, calculated using a MatLab script.
\begin{table}
\centering
\caption{Energy Consumption Contributions}
\label{e_cont}
\setlength{\tabcolsep}{3pt}
\begin{tabular}{lc|lc}
\hline
\hline
\textbf{Contribution}& 
\textbf{Energy [mJ]} &
\textbf{Contribution}& 
\textbf{Energy [mJ]}
\\
\hline
$E_{acq1s}$ & 52.596 &

$E_{SD-WR}$ & 2.1816 \\

$E_{c-1TX}$ & 659.72 &

$E_{pkt-TX}$ & 450.83 \\

$E_{cdrx-disc}$ & 616.97 &
 - & -\\
\hline
\hline
\end{tabular}
\end{table}
The current used by the node in sleep mode was measured using a precise power source meter the Keithley 2470.
When the STM32 enters in Stop Mode 2, all sensors are consequently configured in low power mode, as well as the Quectel BC95-G, which is in PSM. In this configuration, the current absorbed by the board ($I_{sleep}$) is $34~\mu A$. 
The total energy consumed in one day ($E_{day}$) is given by the Eq.~\ref{e_day}, where $V_{s}$ is the supply voltage of 3.3~V, $N_{session}$ is the number of acquisition and transmission performed per day and $T_{sleep}$ is the number of seconds in sleep condition.
\begin{equation}
\footnotesize
    E_{day}=E_{tot}*N_{session}+T_{sleep}*V_{s}*I_{sleep}
    \label{e_day}
\end{equation}
The battery selected for our application is produced by SAFT, named LS336000\footnote{\url{https://www.saftbatteries.com/products-solutions/products/ls-lsh-lsp}}. It is a Lithium - Thionyl chloride ($Li-SOCl_2$) primary cell, it's ideally suited for long-term applications with its 17Ah of capacity in a D-size bobbin cell format. The total energy avaible by this cell is 226440~J.
The battery capacity is temperature dependent. The nominal value is verified at $25^\circ$C, but it could be halved at $-40^\circ$C, thereby modifying results in Fig.~\ref{battery_life}.

\begin{figure}[!t]
\centering
\includegraphics[width=\columnwidth]{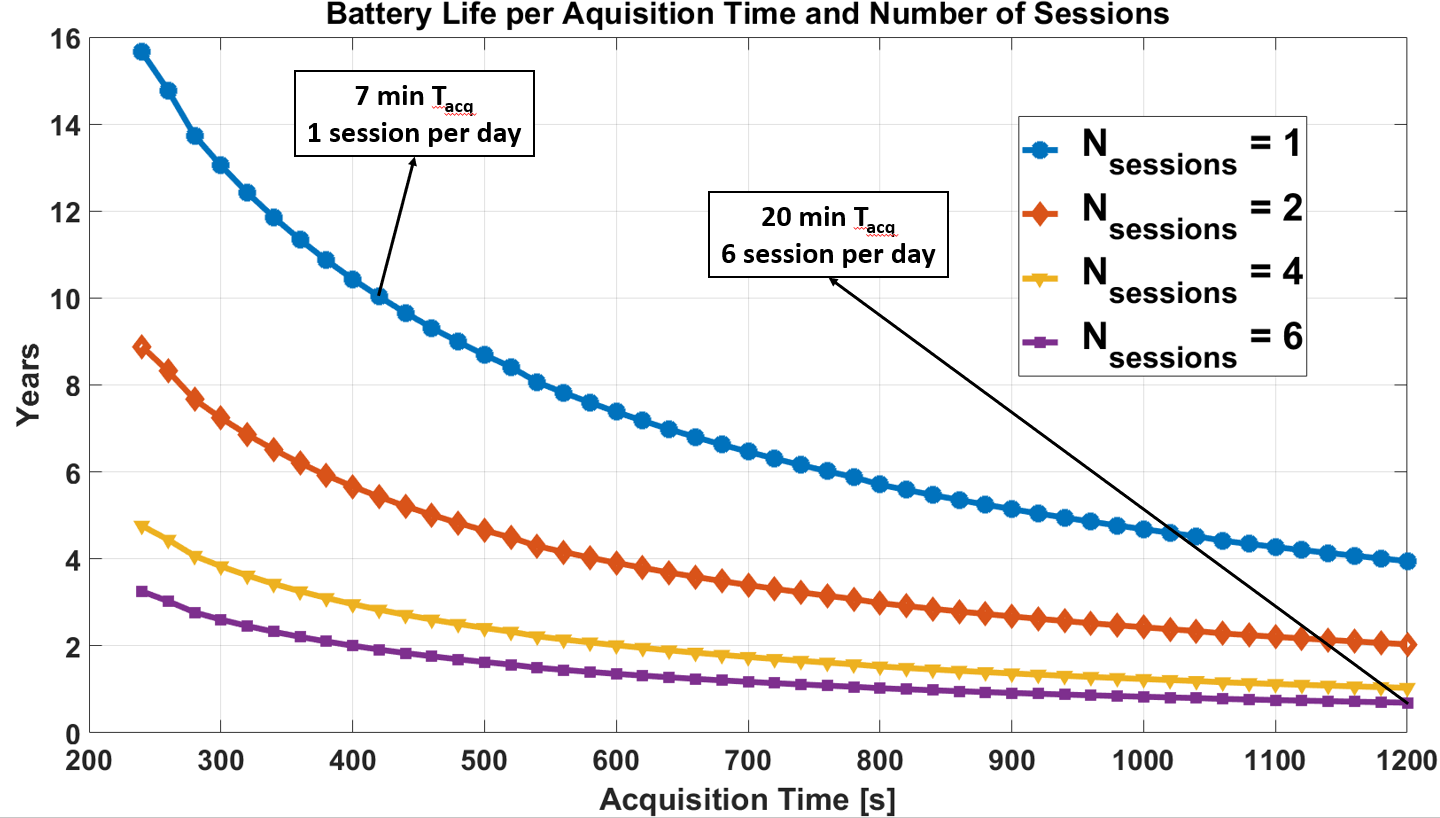}
\caption{Battery life estimation varying  the sensors acquisition time and the number of sessions per day}
\label{battery_life}
\end{figure}
Fig.~\ref{battery_life} shows the battery lifetime (in years) as a function of the acquisition time for each session, and  the number of sessions per day. The acquisition time plotted in Fig.~\ref{battery_life} is from 4 minutes to 20 minutes but it can be lower or higher depending on the application.
We show four different curves according to the different number of sessions per day: 1, 2, 4, 6.
Notice that a configuration with no more than 7 minutes acquisition time and one session per day, is necessary for achieving ten years lifetime, with 84~kB transmitted daily.  
On the other hand, very frequent acquisitions can discharge the battery quite quickly. For example, 6 sessions with 20 minutes of measurement drain the battery charge in 214~days, because 1.44~MB are transmitted every day.

\subsection{Energy Harvesting}
In addtion to aggressive power management, to achieve long term monitoring an energy harvester circuit is designed. Indeed, considering a typical scenario for SHM, where sensors are deployed out in the open on building or bridges, a solar harvesting is a valuable solution. 
In this scenario, a different battery cell is needed because the LS336000 is not rechargeable: a Li-Ion alternative is the VL34570\_xlr
by SAFT. This cell has the same size of the LS336000 (D-size bobbin cell) and the same nominal voltage, $3.7~V$, but the capacity is lower: 5.4~Ah. If we focus on a long term application, we can estimate the energy consumption per day with the aforementioned power model. According to~\cite{SHM_algo}, to perform an analysis of structural damage detection using a time series based algorithm, SHM operators need at least 6000 samples at 100~Hz. To collect this amount of data, our SHM sensor needs 60 seconds of acquisition (from each installed sensor) 6 times per day to check the integrity of the structure.  With this type of battery, using the equations (\ref{num_pkt}), (\ref{acq_e}), (\ref{tx_e}), (\ref{e_tot}), we obtain a duration of 3 years. All results are reported in Table~\ref{shm_e}.
\begin{table}
\centering
\caption{Specs and Data for SHM long term application using solar energy harvesting}
\label{shm_e}
\setlength{\tabcolsep}{3pt}
\begin{tabular}{lc|lc|lc}
\hline\hline
\textbf{Parameter} & \textbf{Value} &
\textbf{Parameter} & \textbf{Value} &
\textbf{Parameter} & \textbf{Value}\\
\hline
$N_{sessions}$ & 6 &

$t_{acq}$ & 60~s &

$t_{active}$ & 546~s \\

$t_{sleep}$ & 85854~s &

$E_{TX}$ & 5.334~J &

$E_{acq}$ & 3.441~J \\

$E_{day}$ & 61.998~J &

$E_{cell}$ & 71928~J &

$Battery Life$ & 3.18~Years \\
\hline\hline
\end{tabular}
\end{table}
The daily energy per day is equal to 61.998~J, with a corresponding average of 17.22~mWh. 

A solar panel in good environment conditions has a power density from 15 to 100~$mW/cm^2$~\cite{Harvesting}. Thus, a solar panel with the same surface as the board (120~mm~$\times$~60~mm) is considered. Supposing a worst case scenario with 4 hours of complete solar light and 25\% loss from recharge and storage circuitry we obtain an average generated power of 3.24~Wh. 
The energy intake from the solar panel each day is 2 orders of magnitude bigger than the power consumed by the node.

\subsection{Model validation}
To check the stability of the software and to validate the power consumption model we extensively tested the sensor node in a real SHM environment. The software is configured to acquire 60 seconds of data and transmit it through NB-IoT protocol to the server, this session is repeated 6 time in a day, one session every 4 hours. 

\begin{figure}[!t]
\centering
\includegraphics[width=\columnwidth]{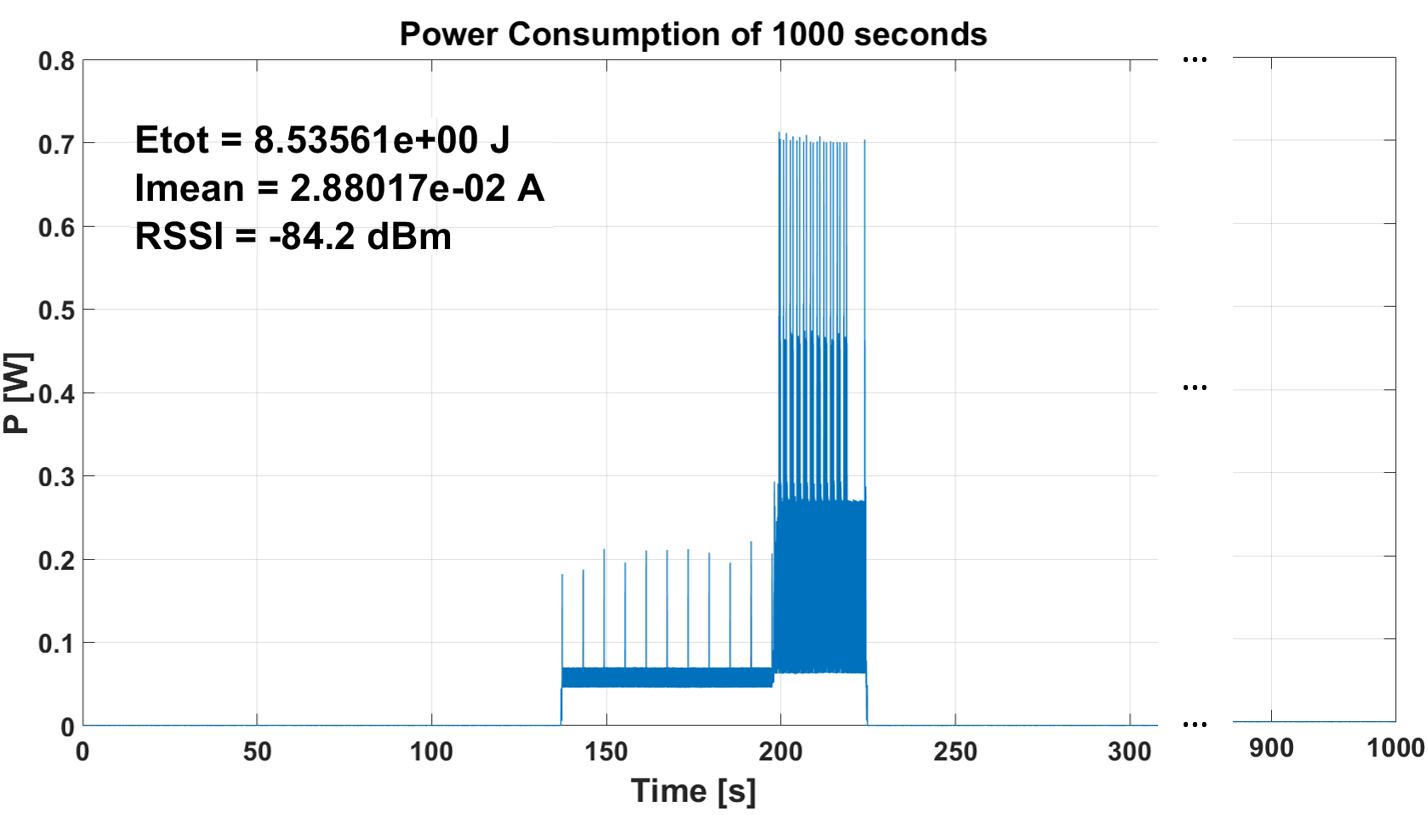}
\caption{Power consumption log of 1000 seconds}
\label{1000sec_acq}
\end{figure} 
In Fig.~\ref{1000sec_acq} we show a log of 1000 seconds, this specific window includes one complete session, in which the total energy consumption is 8.535 J. In the following, a comparison with Equations (\ref{tx_e_model}), (\ref{acq_e_model}), (\ref{e_tot_m}), (\ref{e_sleep_m}), (\ref{e_tot_model}) is proposed in validation of our energy model.
\begin{equation}
\footnotesize
    E_{TX}=E_{c-1TX}+E_{cdrx-disc}+(N_{pkt-TX}-1)E_{pkt-TX}
    \label{tx_e_model}
\end{equation}
\begin{equation}
\footnotesize
     E_{acq}=E_{acq1s}(6*N_{pkt-TX})+N_{pkt-TX}*E_{SD-WR}
    \label{acq_e_model}
\end{equation}
\begin{equation}
\footnotesize
    E_{tot}=E_{TX}+E_{acq} 
    \label{e_tot_m}
\end{equation}
\begin{equation}
\footnotesize
    E_{sleep-1000s}=V_{s}*(1000-T_{sleep})*I_{sleep}
    \label{e_sleep_m}
\end{equation}
\begin{equation}
\footnotesize
    E_{tot-model}=E_{tot}+E_{sleep_1000s}
    \label{e_tot_model}
\end{equation}
From Eq.~(\ref{e_tot_model}), the energy consumed with this window, calculated using our model, is $E_{tot-model}=8.613~J$ that is slightly more than the measured one (0.9\% more). Hence we can conclude that the model overestimates by a small factor the energy consumption and the battery life assumptions are verified. Furthermore the firmware never stopped in 3 days test doing a total of 18 session and transmitting a total of 324~KB to the server. Lastly, the energy consumed in 1 day, measured with our set up, is 62.1~J and considering the SAFT LS336000 cell, we obtain a battery life of 10.1 years.

\section{In-field validation}
\label{sec:in_field_valid}

\begin{figure}[!t]
\centering
\includegraphics[width=\columnwidth]{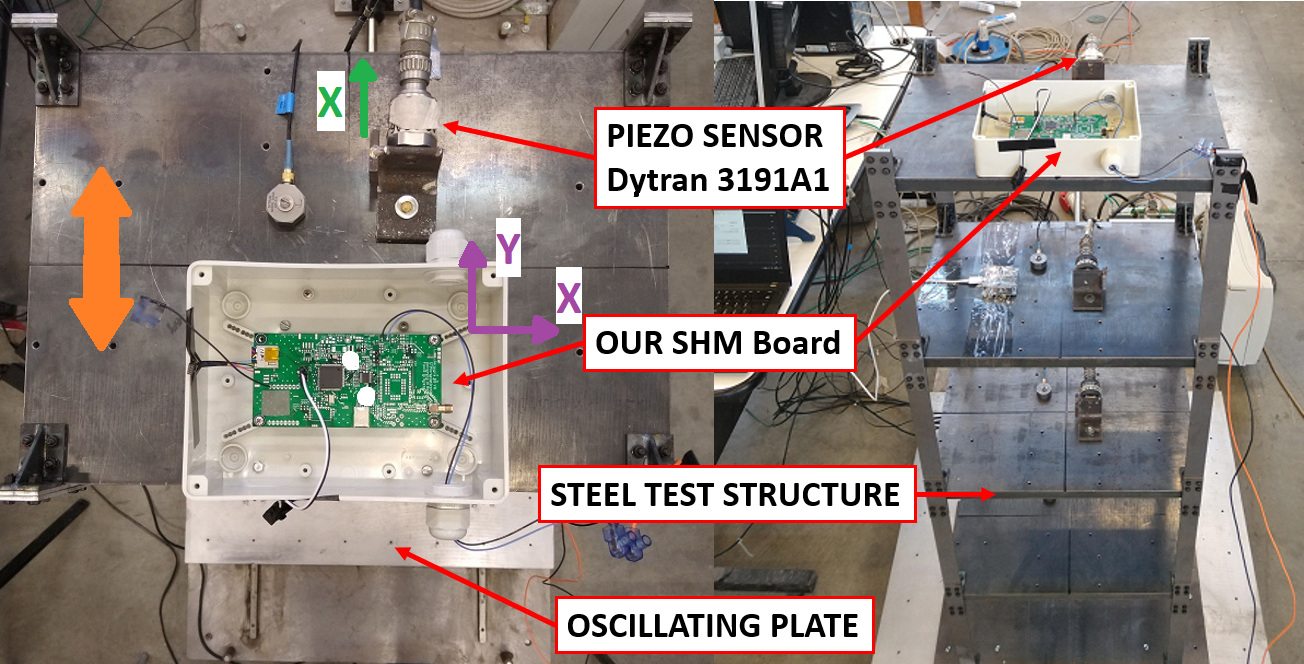}
\caption{
Overview of the test set up: in orange the direction of the oscillations imposed to the structure by the oscillating plate, in violet the axes orientation of our sensor node, and in green the axis orientation of the piezoelectric sensor.
}
\label{set_up}
\end{figure} 

\begin{figure}[!t]
\centering
\includegraphics[width=\columnwidth]{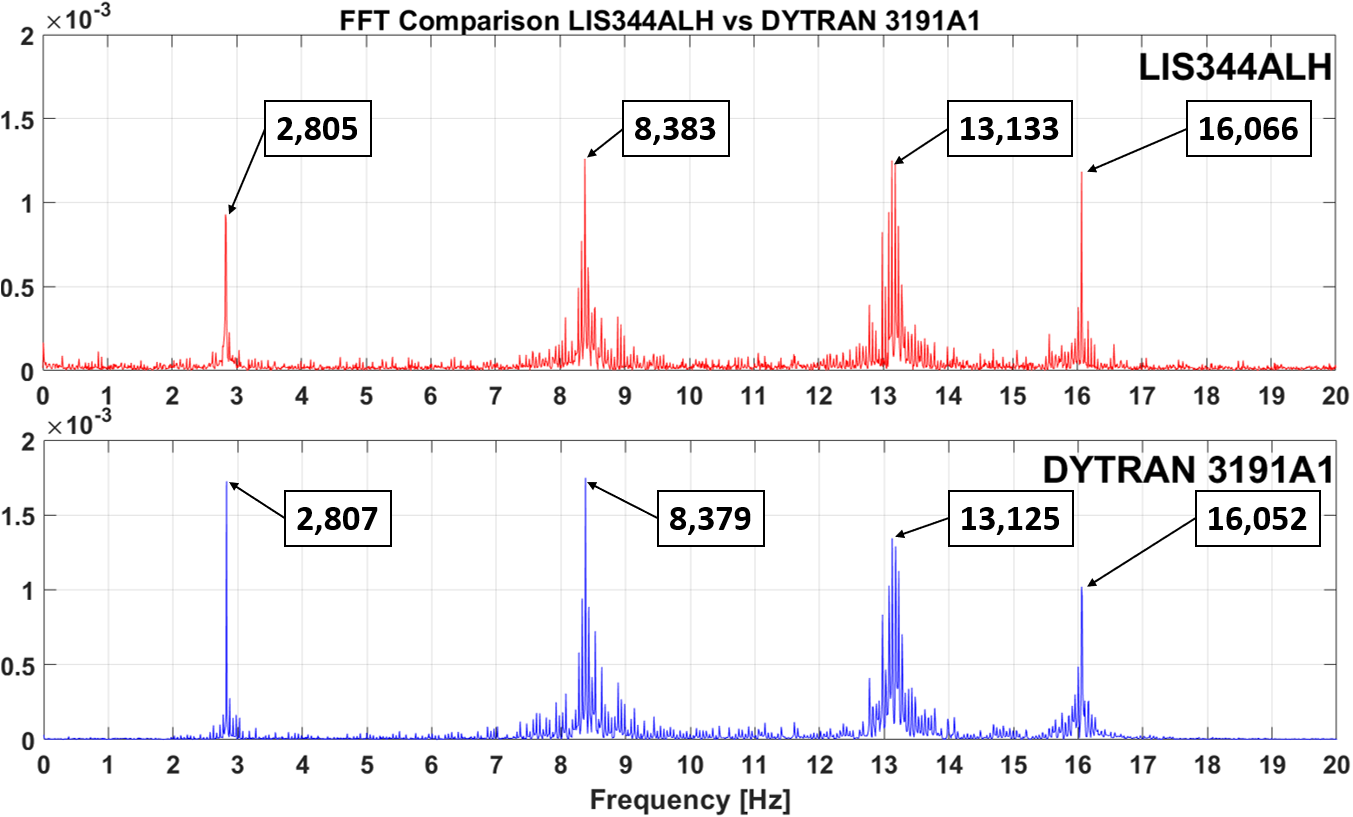}
\caption{FFT analisys of Dytran 3191A1 (bottom graph) and LIS344ALH (top graph) data; the values near tones are the precise frequency.}
\label{dytran_vs_lis}
\end{figure} 

\begin{table}
\centering
\caption{Tone Frequency Comparison LIS344ALH vs Dytran 3191A1}
\label{tone_comp}
\setlength{\tabcolsep}{3pt}
\begin{tabular}{c c c c}
\hline\hline
\textbf{Modal Vibration} & \textbf{LIS344ALH} & \textbf{DYTRAN 3191A1} & \textbf{$\Delta\%$}\\
\hline
I Mode & 2.805 Hz & 2.807 Hz & -0.07\% \\

II Mode & 8.383 Hz & 8.379 Hz & +0.05\% \\

III Mode & 13.133 Hz & 13.125 Hz & +0.06\% \\

IV Mode & 16.066 Hz & 16.052 Hz & +0.08\% \\
\hline\hline
\end{tabular}
\end{table}

\begin{figure}[!t]
\centering
\includegraphics[width=\columnwidth]{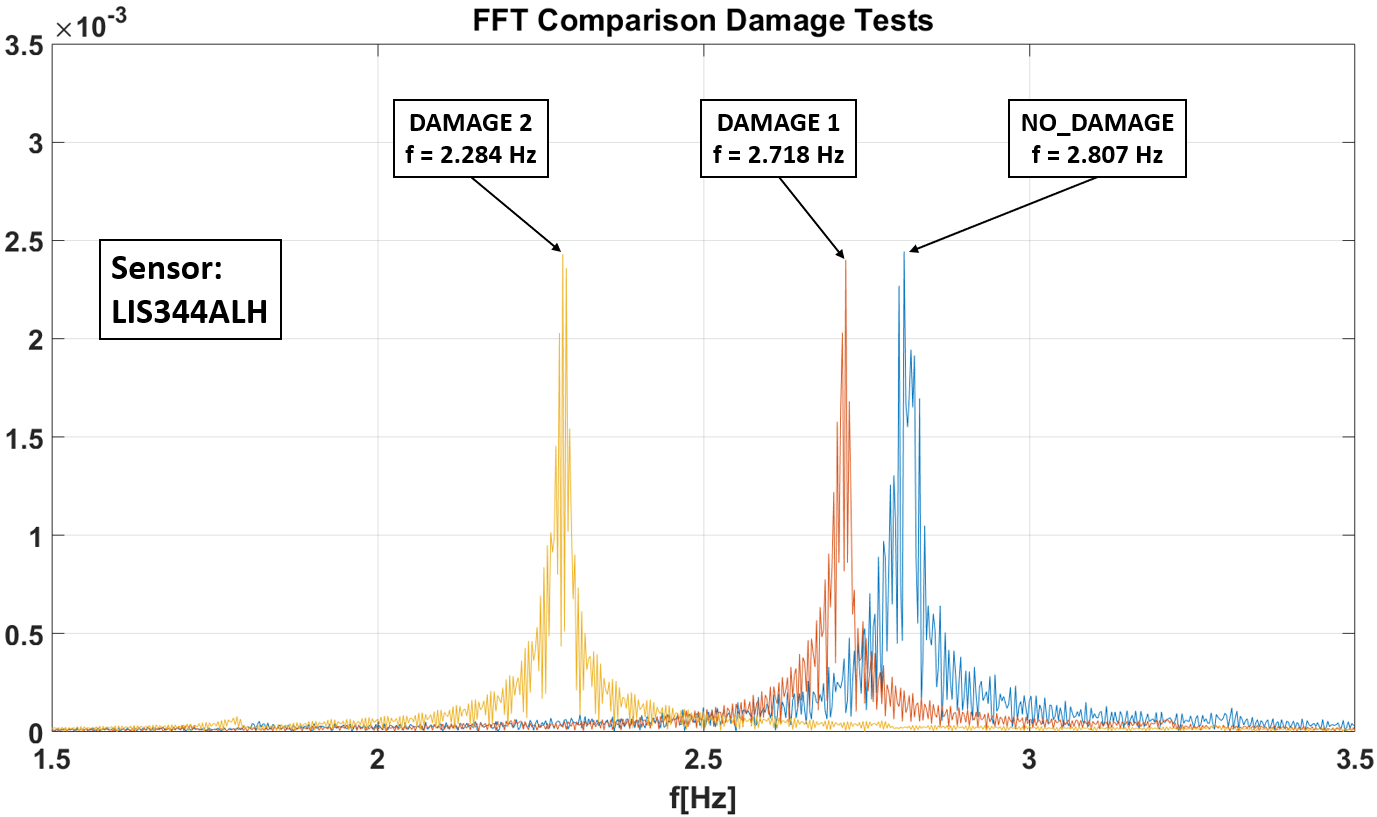}
\caption{FFT analisys of LIS344ALH in 3 test: undamaged structure (NO\_DAMAGE), 
slightly damaged structure (DAMAGE 1) and damaged structure (DAMAGE 2)}
\label{lis_damage}
\end{figure} 

\begin{figure}[!t]
\centering
\includegraphics[width=\columnwidth]{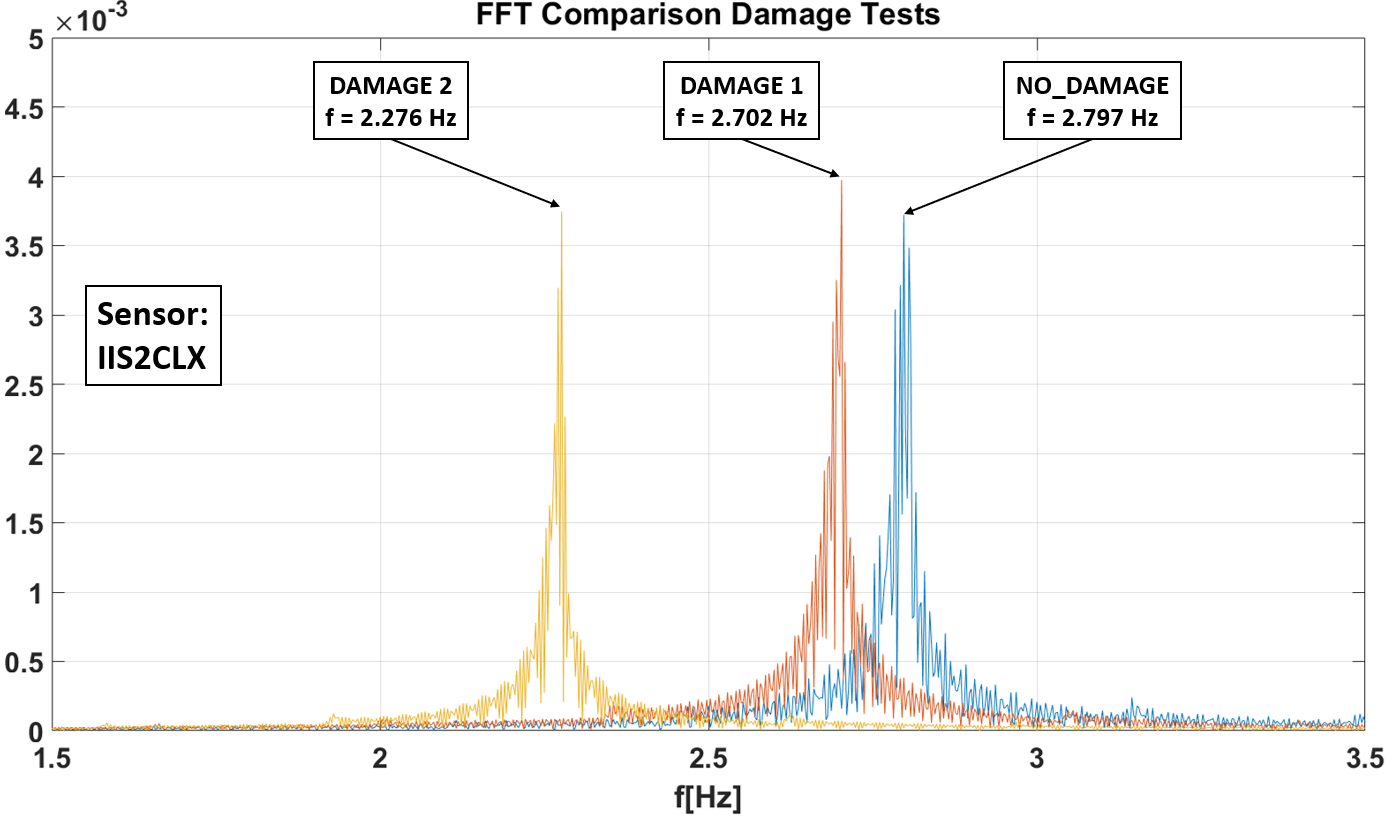}
\caption{FFT analisys of IIS2CLX in 3 test: undamaged structure (NO\_DAMAGE), 
slightly damaged structure (DAMAGE 1) and damaged structure (DAMAGE 2)}
\label{ber_damage}
\end{figure} 

To verify the correctness of acquisitions and to check the measurement accuracy of our SHM sensor node, we assessed the system in a civil engineering test laboratory test structure and environment (Fig.~\ref{set_up}). The steel test structure is
mounted on an oscillating plate that can apply programmed vibration to simulate typical buildings behaviors. The sensor used as reference is a high precision piezoelectric transducer, the Dytran 3191A1\footnote{\url{https://www.dytran.com/Model-3191A-Industrial-Accelerometer-P1625/}}. 
It is specifically designed for SHM applications, and used in high-end instrumentation, as reported in several studies starting from 2010~\cite{baptista210shm}.
A vibration has been imposed on the oscillating plate to see the structure's first four vibration modes on 2.8~Hz, 8.4~Hz, 13.1~Hz, and 16.2~Hz. The data coming from the Dytran 3191A1 and the LIS344ALH has been recorded in multiple measurement sessions of 180~s. 
Fig.~\ref{dytran_vs_lis} shows the two resulting FFTs. Notice that the data from LIS344ALH permit to detect the expected vibration modes, as well as the Dytran 3191A1. Moreover there are no other unexpected frequency components.  The exact frequency values calculated from the two sensors are reported in Table~\ref{tone_comp} and shown in Fig.~\ref{dytran_vs_lis}. The difference is maximum 0.08\%; confirming that Low-cost analog MEMS accelerometers can be used for the assessment of buildings in SHM. 
After the comparison tests, a damage detection simulation has been performed. The steel structure used in the tests (Fig.~\ref{set_up}) can be modified to simulate specific damage, and 2 different situations have been simulated, replacing columns of the steel structure with thinner ones. Fig.\ref{lis_damage} shows the FFT of 3 measurements; the first modal tone is detailed to highlight the frequency shift. Notice that the first modal tone with undamaged structure (Fig.\ref{lis_damage} - NO\_DAMAGE) is at $2.807~Hz$, but when a column of the structure is replaced with a thinner one simulating light damage (Fig.\ref{lis_damage} - DAMAGE 1), the frequency of the first modal tone shifts of $0.089~Hz$ ($2.718~Hz$). In the last test 4 columns of the structure are substituted with thinner ones to simulate moderate damage (Fig.\ref{lis_damage} - DAMAGE 2), and this can be clearly measured in the resulting frequency of the first modal tone $2.284~Hz$ ($0.523~Hz$ less than  ``undamaged'' structure). The same analysis has been executed with data gathered from the IIS2CLX inclinometer sensor. Results show similar modal frequencies, a difference less than 0.3\%. In Fig.\ref{ber_damage} shows the result of the test, with the frequency shift in the 3 cases. 
IIS2CLX is a digital sensor with built-in programmable features like a machine learning core processor and a programmable low pass filter. It is more expensive, with 14.31~€ at 1kunit, it is almost $7\times$ the cost of the LIS344ALH. If the specific SHM application requires machine learning core features, the IIS2CLX could be a viable solution. If the main focus is on cost reduction or the application needs a scale-up to hundreds or thousands of measurement points, the LIS344ALH is the most suitable solution. 
\section{Conclusions}
\label{sec:concl}
Nowadays, the ability to monitor the integrity of a wide variety of buildings with low-cost and real-time measurements is essential from both an economic and a life-safety standpoint. In this work, we proposed a completely untethered wireless sensor node specifically designed to support long-term modal analysis with long-range connectivity at low power consumption.  
We analyzed and tuned the main parameters of the NB-IoT communication protocol, chosen because it provides wireless connectivity to the 4G (and future 5G) global infrastructure network at low power consumption, suitable for continuous monitoring. We provided a comparison of the most recent NB-IoT modules on the market and selected the most appropriate for a continuous SHM scenario. We presented the design, both hardware and software, of a SHM node that can operate unattended for more 10 years, or even indefinitely with the support of a small size solar panel of 72~$cm^2$.
We validated the measurement accuracy of the system by comparing our low-cost devices with a state-of-the-art piezoelectric transducer used in high-end commercial instrumentation for temporary cabled installations. 
Results show a difference lower than 0.08\% in the accuracy of estimation of the modal vibration frequencies, with a cost reduction of around $10\times$. Moreover, our solution enables ease of deployment due to the total absence of cables and long operation lifetime. Lastly, we present three structural damage laboratory tests that confirm the usability of our solution in SHM applications.

%


\begin{thebibliography}{00}

\bibitem{romano2017land}
B.~Romano, F.~Zullo, L.~Fiorini, A.~Marucci, and S.~Ciab{\`o}, ``Land
  transformation of italy due to half a century of urbanization,'' \emph{Land
  use policy}, vol.~67, pp. 387--400, 2017.

\bibitem{calvi2019once}
G.~M. Calvi, M.~Moratti, G.~J. O'Reilly, N.~Scattarreggia, R.~Monteiro,
  D.~Malomo, P.~M. Calvi, and R.~Pinho, ``Once upon a time in italy: The tale
  of the morandi bridge,'' \emph{Structural Engineering International},
  vol.~29, no.~2, pp. 198--217, 2019.

\bibitem{tokognon2017structural}
C.~A. Tokognon, B.~Gao, G.~Y. Tian, and Y.~Yan, ``Structural health monitoring
  framework based on internet of things: A survey,'' \emph{IEEE Internet of
  Things Journal}, vol.~4, no.~3, pp. 619--635, 2017.

\bibitem{shm_zonzini}
F.~{Zonzini}, A.~{Girolami}, L.~{De Marchi}, A.~{Marzani}, and D.~{Brunelli},
  ``Cluster-based vibration analysis of structures with gsp,'' \emph{IEEE
  Transactions on Industrial Electronics}, vol.~68, no.~4, pp. 3465--3474,
  2021.

\bibitem{dhage2017structural}
M.~R. Dhage and S.~Vemuru, ``Structural health monitoring of railway tracks
  using wsn,'' in \emph{2017 International Conference on Computing,
  Communication, Control and Automation (ICCUBEA)}.\hskip 1em plus 0.5em minus
  0.4em\relax IEEE, 2017, pp. 1--5.

\bibitem{valenti2018low}
S.~Valenti, M.~Conti, P.~Pierleoni, L.~Zappelli, A.~Belli, F.~Gara,
  S.~Carbonari, and M.~Regni, ``A low cost wireless sensor node for building
  monitoring,'' in \emph{2018 IEEE Workshop on Environmental, Energy, and
  Structural Monitoring Systems (EESMS)}.\hskip 1em plus 0.5em minus
  0.4em\relax IEEE, 2018, pp. 1--6.

\bibitem{catenazzo2018use}
D.~Catenazzo, B.~O’Flynn, and M.~Walsh, ``On the use of wireless sensor
  networks in preventative maintenance for industry 4.0,'' in \emph{2018 12th
  International Conference on Sensing Technology (ICST)}.\hskip 1em plus 0.5em
  minus 0.4em\relax IEEE, 2018, pp. 256--262.

\bibitem{Burrello_iot}
A.~{Burrello}, A.~{Marchioni}, D.~{Brunelli}, S.~{Benatti}, M.~{Mangia}, and
  L.~{Benini}, ``Embedded streaming principal components analysis for network
  load reduction in structural health monitoring,'' \emph{IEEE Internet of
  Things Journal}, vol.~8, no.~6, pp. 4433--4447, 2021.

\bibitem{haxhibeqiri2018survey}
J.~Haxhibeqiri, E.~De~Poorter, I.~Moerman, and J.~Hoebeke, ``A survey of
  lorawan for iot: From technology to application,'' \emph{Sensors}, vol.~18,
  no.~11, p. 3995, 2018.

\bibitem{ballerini2019experimental}
M.~Ballerini, T.~Polonelli, D.~Brunelli, M.~Magno, and L.~Benini,
  ``Experimental evaluation on {NB-IoT} and lorawan for industrial and iot
  applications,'' in \emph{2019 IEEE 17th International Conference on
  Industrial Informatics (INDIN)}, vol.~1.\hskip 1em plus 0.5em minus
  0.4em\relax IEEE, 2019, pp. 1729--1732.

\bibitem{ballerini2020nb}
M.~Ballerini, T.~Polonelli, D.~Brunelli, M.~Magno, and L.~Benini, ``{NB-IoT}
  vs. lorawan: An experimental evaluation for industrial applications,''
  \emph{IEEE Transactions on Industrial Informatics}, 2020.

\bibitem{sabato2016wireless}
A.~Sabato, C.~Niezrecki, and G.~Fortino, ``Wireless mems-based accelerometer
  sensor boards for structural vibration monitoring: a review,'' \emph{IEEE
  Sensors Journal}, vol.~17, no.~2, pp. 226--235, 2016.

\bibitem{wang2019strain}
H.~Wang, P.~Xiang, and L.~Jiang, ``Strain transfer theory of industrialized
  optical fiber-based sensors in civil engineering: A review on measurement
  accuracy, design and calibration,'' \emph{Sensors and Actuators A: Physical},
  vol. 285, pp. 414--426, 2019.

\bibitem{popkova2019preconditions}
E.~G. Popkova, ``Preconditions of formation and development of industry 4.0 in
  the conditions of knowledge economy,'' in \emph{Industry 4.0: Industrial
  Revolution of the 21st Century}.\hskip 1em plus 0.5em minus 0.4em\relax
  Springer, 2019, pp. 65--72.

\bibitem{shm_brunelli}
A.~{Girolami}, D.~{Brunelli}, and L.~{Benini}, ``Low-cost and distributed
  health monitoring system for critical buildings,'' in \emph{2017 IEEE
  Workshop on Environmental, Energy, and Structural Monitoring Systems
  (EESMS)}, 2017, pp. 1--6.

\bibitem{naeim2007dynamics}
F.~Naeim, ``Dynamics of structures—theory and applications to earthquake
  engineering,'' \emph{Earthquake Spectra}, vol.~23, no.~2, pp. 491--492, 2007.

\bibitem{burrello2020enhancing}
A.~Burrello, D.~Brunelli, M.~Malavisi, and L.~Benini, ``Enhancing structural
  health monitoring with vehicle identification and tracking,'' in \emph{2020
  IEEE International Instrumentation and Measurement Technology Conference
  (I2MTC)}.\hskip 1em plus 0.5em minus 0.4em\relax IEEE, 2020, pp. 1--6.

\bibitem{shm_1999}
C.~R. Farrar, H.~Sohn, and G.~H. Park, ``A statistical pattern recognition
  paradigm for structural health monitoring,'' 01 2004.

\bibitem{shm_damage_detec}
Y.~Liao, A.~Kiremidjian, R.~Rajagopal, and C.-H. Loh, ``Structural damage
  detection and localization with unknown post-damage feature distribution
  using sequential change-point detection method,'' \emph{Journal of Aerospace
  Engineering Vol. 32, Issue 2}, 11 2018.

\bibitem{SHM_algo}
K.~Krishnan~Nair and A.~S. Kiremidjian, ``{Time Series Based Structural Damage
  Detection Algorithm Using Gaussian Mixtures Modeling},'' \emph{Journal of
  Dynamic Systems, Measurement, and Control}, vol. 129, no.~3, pp. 285--293, 08
  2006.

\bibitem{shm_demarchi}
F.~{Zonzini}, M.~M. {Malatesta}, D.~{Bogomolov}, N.~{Testoni}, A.~{Marzani},
  and L.~{De Marchi}, ``Vibration-based shm with upscalable and low-cost sensor
  networks,'' \emph{IEEE Transactions on Instrumentation and Measurement},
  vol.~69, no.~10, pp. 7990--7998, 2020.

\bibitem{polonelli2018crackmeter}
T.~{Polonelli}, D.~{Brunelli}, M.~{Guermandi}, and L.~{Benini}, ``An accurate
  low-cost crackmeter with lorawan communication and energy harvesting
  capability,'' in \emph{2018 IEEE 23rd International Conference on Emerging
  Technologies and Factory Automation (ETFA)}, vol.~1, 2018, pp. 671--676.

\bibitem{wang2018strain2}
H.~Wang, L.~Jiang, and P.~Xiang, ``Improving the durability of the optical
  fiber sensor based on strain transfer analysis,'' \emph{Optical Fiber
  Technology}, vol.~42, pp. 97--104, 2018.

\bibitem{ragam2019shm}
P.~{Ragam} and N.~{Devidas Sahebraoji}, ``Application of mems-based
  accelerometer wireless sensor systems for monitoring of blast-induced ground
  vibration and structural health: a review,'' \emph{IET Wireless Sensor
  Systems}, vol.~9, no.~3, pp. 103--109, 2019.

\bibitem{mekki2019lpwan}
K.~Mekki, E.~Bajic, F.~Chaxel, and F.~Meyer, ``A comparative study of lpwan
  technologies for large-scale iot deployment,'' \emph{ICT Express}, vol.~5,
  pp. 1--7, 03 2019.

\bibitem{alonso2018shmwsnsurvey}
L.~Alonso, J.~Barbarán, J.~Chen, M.~Díaz, L.~Llopis, and B.~Rubio,
  ``Middleware and communication technologies for structural health monitoring
  of critical infrastructures: A survey,'' \emph{Computer Standards \&
  Interfaces}, vol.~56, 09 2018.

\bibitem{arcadius2017shmwsnreview}
C.~{Arcadius Tokognon}, B.~{Gao}, G.~Y. {Tian}, and Y.~{Yan}, ``Structural
  health monitoring framework based on internet of things: A survey,''
  \emph{IEEE Internet of Things Journal}, vol.~4, no.~3, pp. 619--635, 2017.

\bibitem{wanglin}
Y.~.~E. {Wang}, X.~{Lin}, A.~{Adhikary}, A.~{Grovlen}, Y.~{Sui},
  Y.~{Blankenship}, J.~{Bergman}, and H.~S. {Razaghi}, ``A primer on 3gpp
  narrowband internet of things,'' \emph{IEEE Communications Magazine},
  vol.~55, no.~3, pp. 117--123, 2017.

\bibitem{perfNBiot}
A.~{Adhikary}, X.~{Lin}, and Y.~.~E. {Wang}, ``Performance evaluation of nb-iot
  coverage,'' in \emph{2016 IEEE 84th Vehicular Technology Conference
  (VTC-Fall)}, 2016, pp. 1--5.

\bibitem{nb-boundaries}
B.~{Martinez}, F.~{Adelantado}, A.~{Bartoli}, and X.~{Vilajosana}, ``Exploring
  the performance boundaries of nb-iot,'' \emph{IEEE Internet of Things
  Journal}, vol.~6, no.~3, pp. 5702--5712, 2019.

\bibitem{Harvesting}
N.~{Garg} and R.~{Garg}, ``Energy harvesting in iot devices: A survey,'' in
  \emph{2017 International Conference on Intelligent Sustainable Systems
  (ICISS)}, 2017, pp. 127--131.

\bibitem{baptista210shm}
F.~G. {Baptista} and J.~V. {Filho}, ``Optimal frequency range selection for pzt
  transducers in impedance-based shm systems,'' \emph{IEEE Sensors Journal},
  vol.~10, no.~8, pp. 1297--1303, 2010.

\end{thebibliography}

\newpage

\begin{IEEEbiography}[{\includegraphics[width=1in,height=1.25in,clip,keepaspectratio]{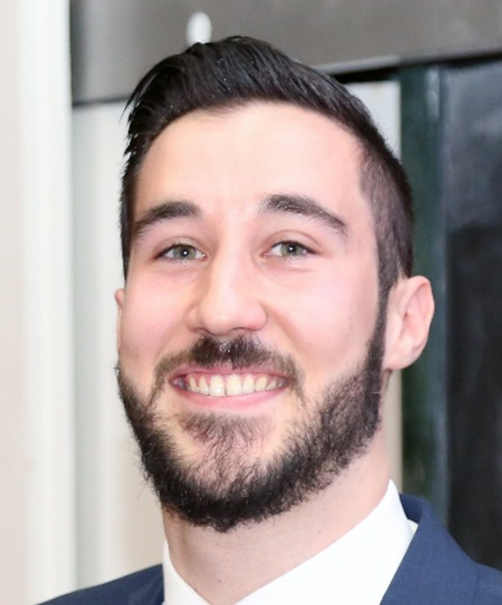}}]{Flavio Di Nuzzo} received the bachelor's degree in electronic engineering and the master's degree in electronic engineering from the University of Bologna in 2016 and 2019, respectively. He has done a year as a research fellow at the University of Bologna studying the Narrowband-IoT technology applied to Civil Structural Health Monitoring. He currently works in Ducati Motor Holding as an Electronic Engineer in the Vehicle Testing department.
\end{IEEEbiography}
\begin{IEEEbiography}[{\includegraphics[width=1in,height=1.25in,clip,keepaspectratio]{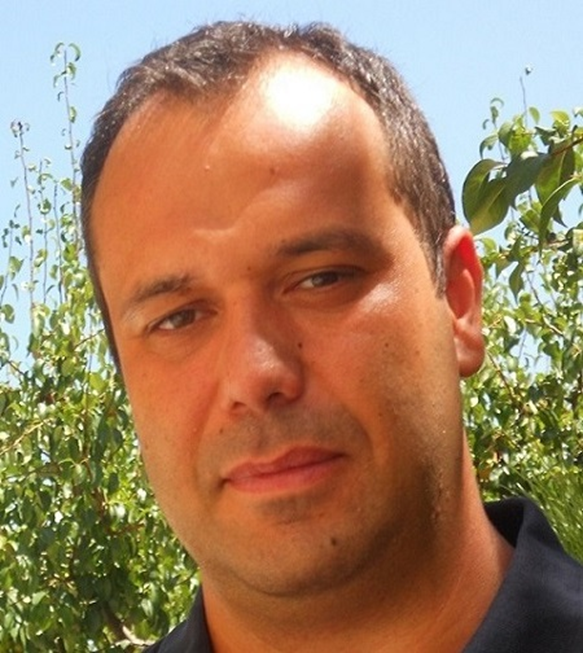}}]{Davide Brunelli} (Senior Member,~IEEE) received the M.S. (cum laude) and Ph.D. degrees in electrical engineering from the University of Bologna, Bologna, Italy, in 2002 and 2007, respectively. He is currently an associate professor at the University of Trento, Italy.  His research interests include IoT and distributed lightweight unmanned aerial vehicles, the development of new techniques of energy scavenging for low-power embedded systems and energy-neutral wearable devices, Drones, UAVs and Machine Learning.  He was leading industrial cooperation activities with Telecom Italia, ENI, and STMicroelectronics. He has published more than 200 papers in international journals or proceedings of international conferences. He is an ACM member.
\end{IEEEbiography}
\begin{IEEEbiography}[{\includegraphics[width=1in,height=1.25in,clip,keepaspectratio]{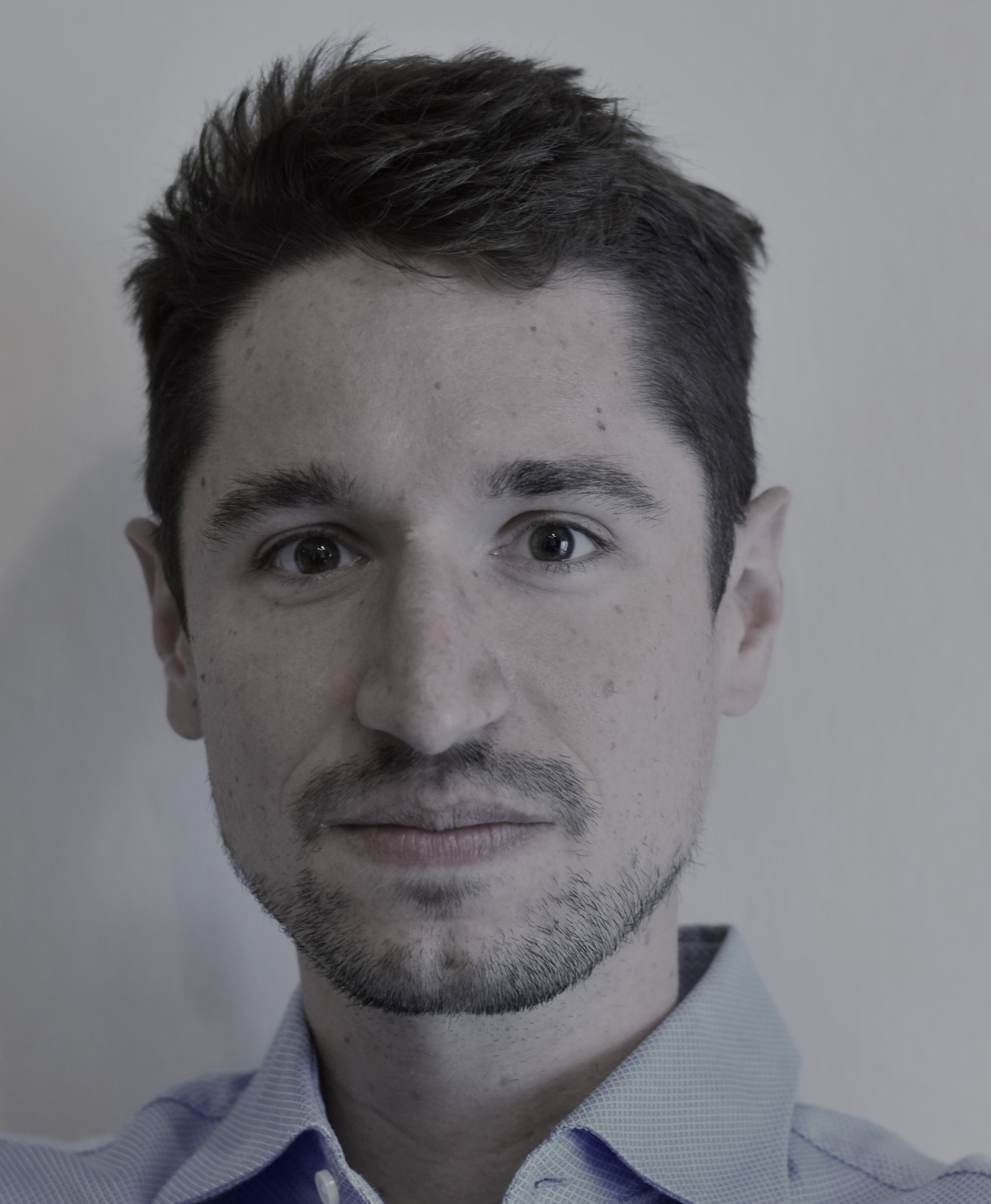}}]{Tommaso Polonelli} (Student Member,~IEEE) received a Ph.D. and a master's degree in electronics and telecommunications engineering from the University of Bologna, Bologna, Italy, in 2020 and 2017, respectively. He is currently a postdoctoral fellow in the center of Project Based Learning, ETH Zurich, Switzerland. His research interests lie in the area of wireless sensor networks, wearable devices, structural health monitoring, autonomous unmanned vehicles, power management techniques, and the design of batteries-operating devices and embedded video surveillance. He has collaborated with several universities and research centers, such as the University College Cork, Cork, Ireland, Imperial College London, London, U.K., and the ETH, Zurich, Switzerland. He has authored over 20 papers in international journals and conferences.
\end{IEEEbiography}
\begin{IEEEbiography}[{\includegraphics[width=1in,height=1.25in,clip,keepaspectratio]{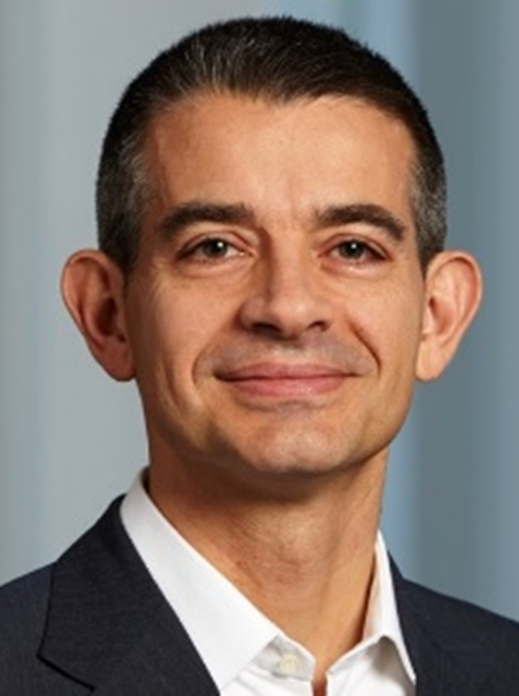}}]{Luca Benini} (Fellow,~IEEE) holds the chair of digital circuits and systems at ETHZ and is Full Professor at the Università di Bologna.
He received a PhD from Stanford University.  He served as chief architect in STMicroelectronics France.
Dr. Benini's research interests are in energy-efficient parallel computing systems, smart sensing micro-systems and machine learning hardware.  
He has published more than 1000 peer-reviewed papers and five books.
He is a fellow of the ACM and a member of the Academia Europaea.
\end{IEEEbiography}

\end{document}